\documentclass{PoS}

\title{ASKAP and MeerKAT surveys\\ of the Magellanic Clouds }

\ShortTitle{ASKAP and MeerKAT surveys of the Magellanic Clouds }

\author{
Jacco Th. van Loon\\
Lennard-Jones Laboratories, Keele University, ST5 5BG, United Kingdom\\
E-mail: \email{jacco@astro.keele.ac.uk}}

\author{The GASKAP Team\\
Hector Arce,
Amanda Bailey,
Indra Bains,
Ayesha Begum,
Kenji Bekki,
Nadya Ben Bekhti,
Joss Bland-Hawthorn,
Kate Brooks,
Chris Brunt,
Michael Burton,
James Caswell,
Maria Cunningham,
John Dickey,
Kevin Douglas,
Simon Ellingsen,
Jayanne English,
Robert Estalella,
Alyson Ford,
Tyler Foster,
Bryan Gaensler,
Jay Gallagher,
Steven Gibson,
Jos\'e Girart,
Paul Goldsmith,
Jos\'e F.\ G\'omez,
Yolanda G\'omez,
Anne Green,
James Green,
Matt Haffner,
Carl Heiles,
Fabian Heitsch,
Patrick Hennebelle,
Tomoya Hirota,
Melvin Hoare,
Hiroshi Imai,
Hideyuki Izumiura,
Gilles Joncas,
Paul Jones,
Peter Kalberla,
Ji-hyun Kang,
Akiko Kawamura,
J\"urgen Kerp,
Charles Kerton,
Bon-Chul Koo,
Roland Kothes,
Stan Kurtz,
Ma\v{s}a Laki\'cevi\'c,
Tom Landecker,
Alex Lazarian,
Nadia Lo,
Felix Lockman,
Loris Magnani,
Naomi McClure-Griffiths,
Karl Menten,
Victor Migenes,
Marc-Antoine Miville-Desch\^enes,
Erik Muller,
Akiharu Nakagawa,
Hiroyuki Nakanishi,
Jun-ichi Nakashima,
David Nidever,
Lou Nigra,
Josh Peek,
Miguel P\'erez-Torres,
Chris Phillips,
Mary Putman,
Anthony Remijan,
Philipp Richter,
Peter Schilke,
Yoshiaki Sofue,
Sne\v{z}ana Stanimirovi\'c,
Daniel Tafoya,
Russ Taylor,
Wen-Wu Tian,
Lucero Uscanga,
Jacco van Loon,
Maxim Voronkov,
Bart Wakker,
Andrew Walsh,
Tobias Westmeier,
Benjamin Winkel,
Ellen Zweibel
}
\author{The MagiKAT Team\\
Gemma Bagheri,
Amanda Bailey,
Sudhanshu Barway,
Yuri Beletsky,
Nadya Ben Bekhti,
Jean-Philippe Bernard,
Michael Bietenholz,
Roy Booth,
Caroline Bot,
Elias Brinks,
Maria-Rosa Cioni,
Maria Cunningham,
Erwin de Blok,
John Dickey,
Kevin Douglas,
Chris Evans,
Brad Gibson,
Steven Gibson,
Sharmila Goedhart,
Jos\'e G\'omez,
Karl Gordon,
Marijke Haverkorn,
Fabian Heitsch,
Patrick Hennebelle,
Benne Holwerda,
Remy Indebetouw,
Frank Israel,
Paul Jones,
Peter Kalberla,
Akiko Kawamura,
J\"urgen Kerp,
B\"arbel Koribalski,
Mark Krumholz,
Ma\v{s}a Laki\'cevi\'c,
Nadia Lo,
Margaret Meixner,
Karl Menten,
Erik Muller,
Joana Oliveira,
Peter Sarre,
Marta Sewi{\l}o,
Lisette Sibbons,
Keith Smith,
Jacco van Loon,
Wouter Vlemmings,
Fabian Walter,
Patricia Whitelock,
Benjamin Winkel,
Albert Zijlstra
}

\abstract{The Magellanic Clouds are a stepping stone from the overwhelming
detail of the Milky Way in which we are immersed, to the global
characteristics of galaxies both in the nearby and distant universe. They are
interacting, gas-rich dwarf galaxies of sub-solar metallicity, not unlike the
building blocks that assembled the large galaxies that dominate groups and
clusters, and representative of the conditions at the height of cosmic star
formation. The Square Kilometre Array (SKA) can make huge strides in
understanding galactic metabolism and the ecological processes that govern
star formation, by observations of the Magellanic Clouds and other, nearby
Magellanic-type irregular galaxies. Two programmes with SKA Pathfinders
attempt to pave the way: the approved Galactic ASKAP Spectral Line Survey
(GASKAP) includes a deep survey in H\,{\sc i} and OH of the Magellanic Clouds,
whilst MagiKAT is proposed to perform more detailed studies of selected
regions within the Magellanic Clouds --- also including Faraday rotation
measurements and observations at higher frequencies. These surveys also close
the gap with the revolutionizing surveys at far-IR wavelengths with the {\it
Spitzer} Space Telescope and {\it Herschel} Space Observatory.}

\FullConference{ISKAF2010 Science Meeting - ISKAF2010\\
		June 10-14, 2010\\
		Assen, the Netherlands}

\begin{document}

\section{The Magellanic Clouds: cosmic evolution up close}

We are incredibly fortuitous to find ourselves in the position to study two
rather substantial actively star-forming dwarf galaxies right on our doorstep
($\approx50$ kpc for the Large Magellanic Cloud and $\approx60$ kpc for the
Small Magellanic Cloud), through a window of transparency. They hold the keys
to unlocking many of the secrets pertaining to the way the Universe has
evolved from what was once a dark place to one of brilliance and activity.
Much of the physical processes at play, within and between galaxies of the
Magellanic type but which are --- or have been --- prevalent in other types of
galactic environment as well, are accessible to the sensitive and sharp radio
eyes of the upcoming pathfinders for the Square Kilometre Array (SKA).

In the cold dark matter hierarchical galaxy assembly paradigm, the young and
dense Universe was teeming with gas clouds whizzing around within the
gravitational lure of nascent clusters of galaxies, arising from fluctuations
in the dark matter distribution. Colliding and merging to form the larger
spiral and elliptical galaxies that are today's Universe's lighthouses, many
of these building blocks have avoided this destiny and are now still the most
plentiful dark-matter dominated systems. Most have since consumed their gas by
converting it into stars and by stripping --- be it by external processes
(galaxy interaction) or by internal processes (winds and supernovae). The more
massive or more isolated among them have been able to retain enough gas to
sustain star formation over prolonged periods of time right till the present
day, leading to the buildup of chemical richesse (the LMC's gas has a metal
content $\approx1/2$ that of the Sun, and the SMC $\approx1/5$). These
gas-rich dwarf galaxies are both tantalising relics of the Universe's re-birth
during the epoch of cosmic re-ionization, as well as fascinating miniature
galaxies in their own right.

The Magellanic Clouds are a splendid showcase of galaxy interaction. The SMC's
reservoir of gas has been depleted by the pull from the significantly more
massive LMC, forming the Magellanic Bridge that connects the two --- star
formation is occurring in the denser portions of this tidal feature. Together,
the Magellanic Clouds tumble through the Milky Way's Halo --- which extends at
least twice as far as where the Magellanic Clouds presently are --- causing
loosely-bound gas to feed a tail that spans half the celestial hemisphere: the
Magellanic Stream. While the tidal forces exerted by the Galaxy become more
effective further down-stream, the ram pressure exerted by the hot Halo gas on
the speedy Magellanic Clouds is tremendous: it has caused the head--tail
morphology of the Clouds--Stream and likely compressed gas at the frontal rim
of the LMC which led to the vigorous formation of stars including the
Tarantula Nebula ``mini-starburst''. Though we probably witness the Magellanic
Clouds during an unusual phase --- their large proper motions cast doubt on
them being bound to the Galaxy (Besla et al.\ 2006) and the SMC is still very
gas-rich --- this is probably not their first (nor last) close encounter with
each other and with the Galaxy.

\section{The Magellanic Clouds: astrophysical laboratories}

The Magellanic Clouds have historically been of great significance for many
fields of astrophysics, thanks largely to their known distance and their
proximity. From the discovery of the period--luminosity relation of Cepheid
variable stars by Henrietta Leavitt in 1908, which set the fundamental
distance scale of the Universe, to the progenitor with the wrong colour of
supernova 1987A, the Magellanic Clouds are full of surprises that turn
long-held beliefs up-side down.

The Magellanic Clouds offer a spectacular manifestation of the multi-phase
interstellar medium (ISM) (Chu 2009): shells produced by fast stellar winds
and supernovae give the warm neutral medium the appearance of bubble-foam;
these are filled with hot plasma, but they are bound to succomb to disruptive
processes including rotational shear, shell--shell collisions, and thermal
instabilities within the shell walls. The dissipation of energy causes a
trickle down the turbulent cascade towards smaller scales, where other sources
await to inject energy and momentum. A question remaining concerns the
importance of magnetic fields in regulating the ISM dynamics.

One of the mysteries in star formation lies at its base: the transformation of
the warm ISM into molecular clouds, and the subsequent collapse of cores on
their way to forming stars. The commonly used tracer of molecular clouds,
carbon-monoxide is detected only in the densest parts, leaving much of the
H$_2$ cloud envelopes unseen --- this may be traced by cold H\,{\sc i} or OH.
Given that the contraction of cores to the point of nuclear fusion has to pass
several hurdles {\it via} cooling by metallic material and {\it via} diffusion
of the lightly-ionized gas through an intensified magnetic field, one might
expect the outcome of this process to depend on metallicity (Oliveira 2009).
Once massive protostars are formed, their outflows and ionizing radiation act
upon the gas surrounding them. Protostars are found at far-IR wavelengths
(Sewi{\l}o et al.\ 2010), by masers, and in later stages as ultra-compact
H\,{\sc ii} regions which shine brightly through free--free emission at radio
wavelengths.

Stars in their final stages lose a significant fraction of their mass in the
form of a stellar wind --- fast for hot stars, slow for cool stars. In the
winds from red (super)giants (except carbon stars), OH maser emission offers
the best way to measure the wind speed, which is connected to the luminosity
and to the gas:dust ratio and hence metallicity; tests in the Magellanic
Clouds are thus extremely useful (Marshall et al.\ 2008). When supernovae and
their remnants expand into the surrounding ISM, they collect matter and
magnetic field from it ({\it e.g.}, Otsuka et al.\ 2010). If they encounter a
molecular cloud they may disrupt it or induce it to collapse --- the outcome
is not yet clear.

\section{Radio surveys of the Magellanic Clouds: a revolution in the making}

The most comprehensive view of the atomic gas in the Magellanic Clouds is had
through the H\,{\sc i} surveys with the Australia Telescope Compact Array and
Parkes dish (LMC: Kim et al.\ 2003; SMC: Stanimirovi\'c et al.\ 1999). These
have an angular resolution of $1^\prime$ (LMC) and $1.6^\prime$ (SMC), a
velocity resolution of 1.6 km s$^{-1}$, and reach brightness sensitivities of
$\sim2$ K. Impressive though they are, they are limited in their ability to
detect cold gas. H\,{\sc i} absorption surveys have detected such gas in a
small number of sightlines through the LMC (Marx-Zimmer et al.\ 2000) and SMC
(Dickey et al.\ 2000). Likewise, Faraday rotation measurements have been made
in several directions suggesting an ordered global field (LMC: Gaensler et
al.\ 2005; SMC: Mao et al.\ 2008). A much denser grid of measurements is
needed to establish the origin of the field and the r\^ole it plays at the
scales of star-forming regions in the Magellanic Clouds (1--$10^\prime$).

Radio continuum emission traces free electrons {\it via} free--free emission
or synchrotron emission if gyrating within a magnetic field. Surveys of the
Magellanic Clouds have been conducted at frequencies between 1.4--8.6 GHz
(LMC: Dickel et al.\ 2005; Hughes et al.\ 2007; SMC: Filipovi\'c et al.\ 2002)
at scales between 20--$40^{\prime\prime}$ --- smaller for bright sources such
as supernova remnants.

Masers have been detected in the LMC from protostars (Ellingsen et al.\ 2010;
{\it cf.}\ Oliveira et al.\ 2006) and evolved stars (Marshall et al.\ 2004;
{\it cf.}\ van Loon et al.\ 2001), but not in the SMC.

Whilst existing H\,{\sc i} surveys are a good match to the far-IR resolution
of IRAS, modern IR surveys have progressed to much greater detail: around
6--$40^{\prime\prime}$ at 24--160 $\mu$m with the {\it Spitzer} Space
Telescope and a similar range but at 60--600 $\mu$m with the {\it Herschel}
Space Observatory (Meixner et al.\ 2010). This means that currently the
resolution of pseudo-H$_2$ maps that can be derived from the combination of
H\,{\sc i} and far-IR data is (severely) limited by that of the H\,{\sc i}
data.

Planned and proposed surveys of the Magellanic Clouds with Southern SKA
pathfinders, to start in 2013, GASKAP and MagiKAT will be deeper and sharper
in both an angular and kinematic sense, by an order of magnitude or more over
existing surveys (Table 1).

\begin{table}[h]
\begin{tabular}{|l|l|l|}
\hline
Survey & Tracer & Particulars \\
\hline
{\it GASKAP}  &
 H\,{\sc i} emission/absorption &
 $S_{\rm rms}=0.76$ K at $20^{\prime\prime}$, 1 km s$^{-1}$
 (also $8^{\prime\prime}$, 0.2 km s$^{-1}$) \\
 & OH emission/absorption         &
 few mJy ($5\sigma$), at 1612+1665+1667 MHz \\
\hline
{\it MagiKAT}
 & H\,{\sc i} emission/absorption &
 $S_{\rm rms}=0.49$ K at $20^{\prime\prime}$, 1 km s$^{-1}$
 (also $8^{\prime\prime}$, 0.2 km s$^{-1}$) \\
 & OH emission/absorption         &
 few mJy ($5\sigma$), at 1612+1665+1667+1720 MHz \\
 & CH$_3$OH masers                &
 at 12.2 GHz, few mJy ($5\sigma$), few pointings per $1^\circ$ field \\
 & high-frequency continuum       &
 at 8--14 GHz, full Stokes, $\sim1^{\prime\prime}$ resolution \\
 & Faraday rotation               &
 at 0.6--2.5 GHz, $\sim1000$ measurements per $1^\circ$ field \\
\hline
\end{tabular}
\caption{Summary of the approved GASKAP and proposed MagiKAT surveys of the
Magellanic Clouds.}
\end{table}

\subsection{GASKAP}

The Galactic ASKAP Spectral Line Survey (GASKAP) is an approved Survey Science
Project at the Australian SKA Pathfinder (ASKAP). It includes the Magellanic
Clouds among its deepest survey components (yet a modest 627 hr) in H\,{\sc i}
and OH 1612, 1665 and 1667 MHz. The resolution is nominally
$20^{\prime\prime}$ but postage-stamp cubes are planned for the brightest
compact sources at $8^{\prime\prime}$.

\begin{figure}[b]
\begin{center}
\includegraphics[width=0.94\textwidth]{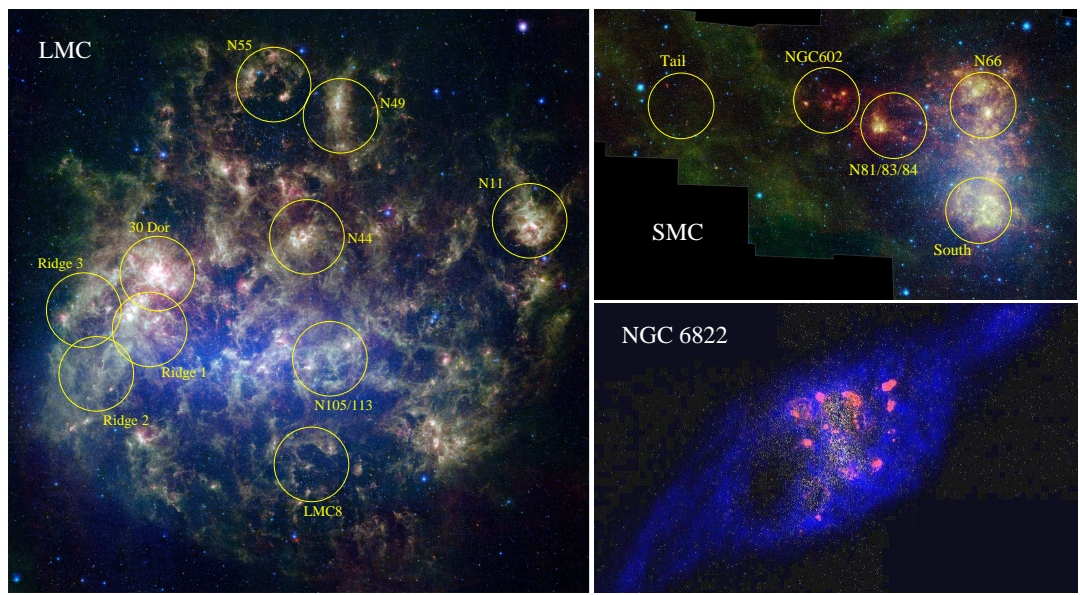}
\vspace*{-5mm}
\end{center}
\caption{Proposed target fields for MagiKAT --- circles are for 1.4-GHz
($1^\circ$), NGC\,682 is covered entirely.}
\end{figure}

\subsection{MagiKAT}

MagiKAT is a proposed Large Project for the South African SKA pathfinder,
MeerKAT. It specifically targets the Magellanic Clouds ($\approx4400$ hr, plus
the next-nearest Magellanic-type galaxy, NGC\,6822), but only in ten selected
fields in the LMC and five in the SMC, each $1^\circ$ in diameter at 1.4 GHz
(Fig.\ 1). Besides H\,{\sc i} and all four OH lines it also measures the
continuum at higher and searches for methanol masers at 12.2 GHz --- in
correspondingly smaller regions within the $1^\circ$ fields --- and it will
obtain Faraday rotation measurements of the magnetic field by observing
background radio galaxies between 0.6--2.5 GHz. The expectation is that over
$10^4$ Faraday rotation measurements will be performed, along with hundreds
H\,{\sc i} absorption measurements.

\section{Gearing up for the future}

\begin{figure}[h]
\begin{center}
\includegraphics[width=0.52\textwidth]{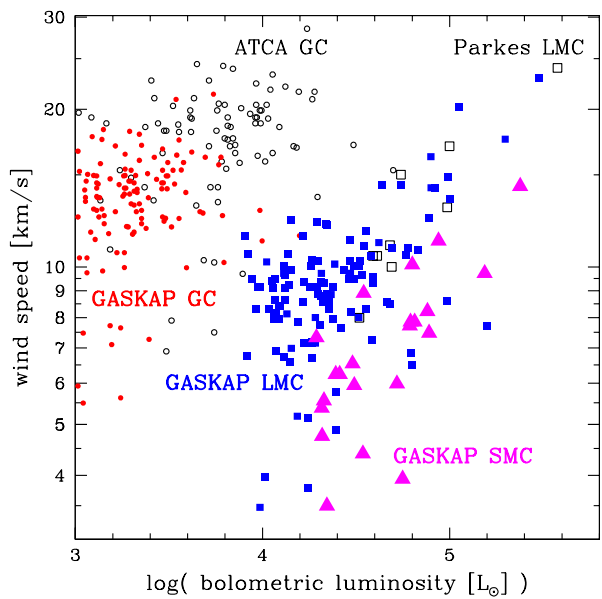}
\vspace*{-5mm}
\end{center}
\caption{Prediction for the yield of OH 1612-MHz masers from evolved stars in
the GASKAP survey.}
\end{figure}

GASKAP has entered the design phase. As part of this endeavour, it has just
completed an H\,{\sc i} simulation through the ASKAP simulator and pipeline on
scaled data from the Galactic All-Sky Survey (GASS). It is now finalising
preparations for a realistic OH maser simulation, improving upon the scaling
of known Galactic populations (Fig.\ 2). These simulations are used to assess
data reduction requirements, and to test source finding algorithms; in the
case of the OH maser simulations these can be compared with real observations
with ASKAP (and MeerKAT) to test the validity of the underlying assumptions
regarding the OH maser populations.

Exciting times lie ahead for these two superb pathfinders on the road to
unlocking the secrets of the Magellanic Clouds, one giant step towards
understanding the workings of the Universe. They will also demonstrate what
further can be gained from using the SKA for studies of diffuse matter.

\acknowledgments
The author cherishes happy memories of a lovely time with charming housemates.

\end{document}